\documentclass[aps,twocolumn,showpacs,preprintnumbers,amsmath,amssymb,nofootinbib,superscriptaddress,showkeys]{revtex4}

\usepackage{hyperref}
\usepackage{float}
\usepackage{epsfig}
\usepackage{graphicx}
\usepackage{mathptmx}      
\usepackage{latexsym}
\usepackage{longtable}
\usepackage{dcolumn}

\begin{document}

\title{Triton binding energy with realistic
  precision}

\author{R. Navarro P\'erez}\email{rnavarrop@ugr.es}
\affiliation{Departamento de F\'{\i}sica At\'omica, Molecular y
  Nuclear \\ and Instituto Carlos I de F{\'\i}sica Te\'orica y
Computacional \\ Universidad de Granada, E-18071 Granada, Spain.}
\author{E.Garrido}\email{e.garrido@csic.es}
\affiliation{Instituto de Estructura de la Materia, CSIC, Madrid, Spain.}
\author{J.E. Amaro}\email{amaro@ugr.es} \affiliation{Departamento de
  F\'{\i}sica At\'omica, Molecular y Nuclear \\ and Instituto Carlos I
  de F{\'\i}sica Te\'orica y Computacional \\ Universidad de Granada,
  E-18071 Granada, Spain.}  \author{E. Ruiz
  Arriola}\email{earriola@ugr.es} \affiliation{Departamento de
  F\'{\i}sica At\'omica, Molecular y Nuclear \\ and Instituto Carlos I
  de F{\'\i}sica Te\'orica y Computacional \\ Universidad de Granada,
  E-18071 Granada, Spain.} 
\date{\today}

\begin{abstract} 
\rule{0ex}{3ex} We compute the binding energy of triton with realistic
statistical errors stemming from NN scattering data uncertainties and
the deuteron and obtain $E_t=-7.638(15) \, {\rm MeV}$. Setting the
numerical precision as $\Delta E_t^{\rm num} \lesssim 1 \, {\rm keV}$ we
obtain the statistical error $\Delta E_t^{\rm stat}= 15(1) \, {\rm keV}$
which is mainly determined by the channels involving relative
S-waves. This figure reflects the uncertainty of the input NN data, more than two orders of magnitude larger than the experimental
precision $\Delta E_t^{\rm exp}= 0.1 \, {\rm keV}$ and provides a
bottleneck in the realistic precision that can be reached. This
suggests an important reduction in the  numerical precision
and hence in the computational effort. 
\end{abstract}

\maketitle

One of the main challenging goals in theoretical Nuclear Physics is
the {\it ab initio} determination of binding energies of atomic
nuclei. The accepted protocol consists of undertaking a quantum
multinucleon calculation from the knowledge of few-body forces. The
simplest case were such a program has been most often investigated is
the binding energy of triton, a stable system consisting of two
neutrons and a proton with an experimental mass deffect given
currently by $M_t - 2 m_n - m_p = E_t^{\rm exp}= -B_t^{\rm exp} =
-8.4820(1) \, {\rm MeV}$. Already in the mid 1930's quantum mechanical
theoretical studies of triton binding allowed to establish essential
properties of the nuclear force: its finite range as well as the
existence of neutron-neutron interactions (see
e.g. Refs.~\cite{Bethe:1936zz,blatt19521952theoretical} for early
reviews). The increasing precision in our knowledge of the two body
interaction has strongly motivated the developments in solving the
computationally expensive 3N problem
(see
e.g. \cite{amado1969three,delves1969present,wu1993three,glockle1983quantum}).
While this was partly aimed at establishing the need of 3N forces,
high numerical precision in conjunction with realistic and precise
nucleon-nucleon interactions has become a major issue by itself in
few-body computational methods. In
Refs.~\cite{Chen:1985zzb,Elster:1998qv,kievsky1997high} benchmarking
precisions of $\Delta E_t^{\rm num}=10,0.1,0.01 \, {\rm keV}$ have been
achieved within different schemes.

However, nucleon-nucleon potentials determined from data inherit
statistical fluctuations that propagate to the triton theoretical
energy into a genuine statistical error $\Delta E_t^{\rm stat}$. A
pioneering and forgotten attempt already looked at the consequences
for triton binding based on an analysis of the inverse scattering in
the $^1S_0$ channel~\cite{adam1993error}. In the present paper we
quantify for the first time the uncertainty of triton energy $\Delta
E_t^{\rm stat}$ stemming from a complete statistical analysis of 6713
selected nucleon-nucleon scattering data.

The main and most reliable source of information for the NN
interaction are the deuteron energy and the more than 8000 np and pp
scattering data below pion production threshold published during the
last 65 years. These will be denoted as $O_i^ {\rm exp} \pm \Delta
O_i$, with $i=1, \dots, N$ and will be regarded as normally
distributed variables. In the classical statistical approaches one
proposes a given NN interaction $V_{NN}({\bf p})$ dependending on a
set of parameters ${\bf p}=(p_1, \dots,p_P)$ which, by solving the two
body Schr\"odinger equation, generates a set of scattering observables
$O_i ({\bf p})$ with $i=1, \dots , N$.  The parameters are
determined by a least squares $\chi^2$-fit,
\begin{eqnarray}
\min_{\bf p} \chi^2({\bf p}) = 
\min_{\bf p} 
\sum_{i=1}^{N} 
\left( 
\frac{O_i^{\rm exp}-O_i({\bf p})}{\Delta O_i^{\rm exp}}
\right)^2 
\equiv 
\chi^2 ({\bf p}_0) \, .   
\label{eq:chi2}
\end{eqnarray}
A {\it high quality} potential is one verifying $\chi^2/\nu \sim 1 $,
with $\nu=N-P$.  Since the Nijmegen group analysis   in
1993~\cite{Stoks:1993tb} a set of high quality potentials have emerged
fitting their contemporary
databases~\cite{Stoks:1994wp,Wiringa:1994wb,Machleidt:2000ge,Gross:2008ps,NavarroPerez:2012qf,Ekstrom:2013kea,Perez:2013mwa,Perez:2013jpa,Perez:2013oba,Perez:2014yla}. However,
the self-consistency of the $\chi^2$ approach requires the residuals
to be normally distributed,
\begin{equation}
 R_i = \frac{O_i^{\rm exp} - O_i({\bf p}_0)}{\Delta O_i^{\rm exp}} \sim N(0,1),
 \label{eq:normalresiduals}
\end{equation}
a condition which, even if elementary, has only recenty been
addressed~\cite{Perez:2014yla} and checked in the previous
analyses~\cite{Perez:2013mwa,Perez:2013jpa,Perez:2013oba}. The total
number of np and pp data was $N=6713$. This is almost twice as in the
1993 Nijmegen analysis~\cite{Stoks:1993tb} that lacked a normality
test. The normality property of the residuals has been exploited to
extract the effective interaction parameters and corresponding
counterterms~\cite{Perez:2014kpa} and to replicate via Monte Carlo
bootstrap simulation as a means to gather more robust information on
the uncertainty characteristics of fitting
parameters~\cite{Perez:2014jsa}.  We stress that the verification of
normality, Eq.~(\ref{eq:normalresiduals}), is essential for a
meaningful propagation of the statistical error, since the uncertainty
inherited from the fitted scattering data $\Delta O_i^{\rm exp}$
corresponds to a genuine statistical fluctuation.  This allows to
determine the $1\sigma$ error of the parameters ${\bf p}= {\bf p}_0
\pm \Delta {\bf p}^{\rm stat}$ and hence the error in the potential
\begin{equation}
V_{NN} = V_{NN} ( {\bf p}_0 ) \pm \Delta V_{NN}^{\rm stat} 
 \label{eq:potential-we}
\end{equation}
which generates in turn the error in the NN phase-shifs $\delta =
\delta ({\bf p_0}) \pm \Delta \delta^{\rm stat}$ and mixing
angles. Once the NN-potential is determined the three body problem can be solved for the triton binding energy, 
\begin{equation}
\left[\sum_i T_i + \sum_{i<j} V_{NN}(ij) \right] \Psi = E_t \Psi
 \label{eq:triton}
\end{equation}
where 
\begin{equation}
E_t = E_t ({\bf p}_0) \pm \Delta E_t^{\rm stat} \, . 
\end{equation}

Direct methods to determine $\Delta {\bf p}^{\rm stat}$, $\Delta
V_{NN}^{\rm stat}$ and $\Delta E_t^{\rm stat} $ proceed either by the
standard error matrix or Monte Carlo methods (see
e.g.~\cite{Nieves:1999zb}). In Ref.~\cite{Perez:2014jsa} we have shown
that the latter method is more convenient for large number of fiting
parameters (typically $P=40-60$), and consists of generating a
sufficiently large sample drawn from a multivariate normal probability
distribution
\begin{equation}
\label{eq:multinormal}
 P(p_1,p_2,\ldots,p_P) = \frac{1}{\sqrt{(2 \pi)^P \det {\cal E}}}
 e^{-\frac{1}{2}({\bf p}- {\bf p}_0)^T {\cal E}^{-1} ({\bf p}- {\bf
 p}_0)},
\end{equation}
where ${\cal E}_{ij}= (\partial^2 \chi / \partial p_i \partial
p_j)^{-1}$ is the error matrix. We generate $M$ samples ${\bf
  p}_\alpha \in P $ with $\alpha=1, \dots , M$, and compute $V_{NN}(
{\bf p_\alpha})$ from which the corresponding scattering phase shifts 
$\delta ({\bf p}_\alpha)$ and triton binding energies $E_t ( {\bf
  p_\alpha})$ can be determined.

In our calculations we take $M=205$ samples for the smooth potential
described in \cite{Perez:2014yla} ($r_c = 3 \, {\rm fm})$,
\begin{eqnarray}
   V(\vec r) = V_{\rm short} (r) \theta(r_c-r)+ V_{\rm long} (r) \theta(r-r_c)\, . 
\label{eq:potential}
\end{eqnarray}
The long-range piece $V_{\rm long}(\vec r)$ contains a
charge-dependent (CD) one pion exchange (OPE) with fixed
$f^2=0.075$~\cite{Stoks:1992ja}) and electromagnetic (EM) corrections
which are kept fixed throughout the fitting process. The short-range
component is
\begin{eqnarray}
   V_{\rm short}(\vec r) = \sum_{n=1}^{21} \hat O_n \left[\sum_{i=1}^N V_{i,n} e^{-r^2/ (2a_i^2)} \right]
\, , 
\label{eq:potential-gauss}
\end{eqnarray}
where $ \hat O_n$ are the set of operators in the extended AV18
basis~\cite{Wiringa:1994wb,NavarroPerez:2012vr,Perez:2012kt,Amaro:2013zka},
$V_{i,n}$ are fitting parameters and $a_i=a/ (i+1)$ with $a =2.3035
\pm 0.0133 \, {\rm fm}$. For this potential $\chi^2/\nu = 1.06$ and
normality of residuals is verified. The potential uncertainties
$\Delta V_{NN}^{\rm stat}$ have been depicted in~\cite{Perez:2014yla}.
We have checked that statistical uncertainties in the phases and
mixing angles $\Delta \delta^{\rm stat} $ determined by the covariance
matrix method (which would correspond to the limit $M \to \infty$) are
fairly well reproduced by our $M=205$ samples when the variance of the
population is used as an estimator. Likewise, the uncertainties of the
potential Eq.~(\ref{eq:potential-gauss}) obtained by the multivariate
distribution, Eq.~(\ref{eq:multinormal}) are in fair agreement with
our original partial wave analysis to the $3 \sigma$ self consistent
database in terms of a delta-shell potential with OPE
(DS--OPE)~\cite{Perez:2013jpa} and also with the corresponding
bootstrap simulation~\cite{Perez:2014jsa}.

The results for $B_t$ for each one ot the $M=205$ Monte Carlo samples
of the potential have been obtained by means of the Hyperspherical
Adiabatic Expansion Method described in \cite{Nielsen:2001}.  The
angular part of the Faddeev equations is first solved for fixed values
of the hyperradius $\rho$. The corresponding angular eigenfunctions
$\{\Phi_n(\rho, \Omega)\}$ form a complete set, and it is used as a
basis in order to expand the total three-body wave function $\Psi$ as
\begin{equation}
\Psi=\frac{1}{\rho^{5/2}}\sum_n f_n(\rho) \Phi_n(\rho,\Omega),
\label{3bd}
\end{equation}
where $\Omega$ collects the usual five hyperangles, and where the
radial wave functions $f_n(\rho)$ are obtained in a second step by
solving a coupled set of differential radial equations where the
eigenvalues of the angular part enter as effective potentials (see
Ref.\cite{Nielsen:2001} for details).

When solving the angular part, the eigenfunctions
$\Phi_n(\rho,\Omega)$ are expanded in terms of the Hyperspherical
Harmonics (HH), which contain the dependence on the quantum numbers
$\{\ell_x, \ell_y, L,s_x,s_y,S\}$ of the different components included
in the calculation. Obviously, $\ell_x$ and $s_x$ are the relative
orbital angular momentum and spin of one of the two-body subsystems in
the triton, $\ell_y$ is the relative orbital angular momentum between
the third particle and the center of mass of the two-body system, and
$s_y$ is the spin of the third particle. The angular momenta $\ell_x$
and $\ell_y$ couple to $L$, and $s_x$ and $s_y$ couple to the total
spin $S$. Finally, $L$ and $S$ couple to the total angular momentum
1/2 of the triton ground state. Together with these quantum numbers
the HH depend of the hypermomentum $K=2\nu+\ell_x+\ell_y$
($\nu=0,1,2,\cdots$).

Therefore, the convergence of the three-body wave function $\Psi$ has
to be achieved at three different levels. First, in terms of the
adiabatic channels included in the expansion explicitly written in
Eq.(\ref{3bd}). Second, in terms of the components (with quantum
numbers $\{\ell_x, \ell_y, L,s_x,s_y,S\}$) included in the expansion
of the angular functions $\{\Phi_n\}$.  And third, in terms of the
maximum value of the hypermomentum, $K_{max}$, used for each of the
components.  In the calculations presented here we have included up to
12 adiabatic terms in the expansion in Eq.(\ref{3bd}) (typically, four
or five terms are enough to get a good convergence for bound
states). All the partial waves with $\ell_x, \ell_y \leq 5$ have been
included (when increasing the number of components to $\ell_x, \ell_y
\leq 8$ no substantial difference has been observed). Finally, three
different sets of $K_{max}$-values have been considered.  We shall
refer to them as sets $(i)$, $(ii)$, and $(iii)$.  In set $(i)$, about
500 HH are used in total, and $K_{max}=50$ for the most relevant
component in the three-body wave function (which corresponds to
$\ell_x=0$ and $s_x=1$ between the proton and one of the neutrons, and
$\ell_y=0$). In set $(ii)$ we multiply all the $K_{max}$-values by 2
(which means about 1000 HH in the three-body wave function and
$K_{max}=100$ for the dominating component). Finally, in set $(iii)$
we again multiply all the $K_{max}$-values by 2 (therefore, about 2000
HH in the three-body wave function and $K_{max}=200$ for the
dominating component).  An appropriate choice of the $K_{max}$-values
is crucial in order to optimize the computing time. An increase of the
total number of HH in the calculation by a certain factor implies an
increase of the computing time of basically the same factor
squared. As an example, while a single three-body calculation with set
$(i)$ lasts for about 30 minutes, the same calculation with set
$(iii)$ requires no less than 8 hours.

\begin{figure}
\centering
\includegraphics[width=\linewidth]{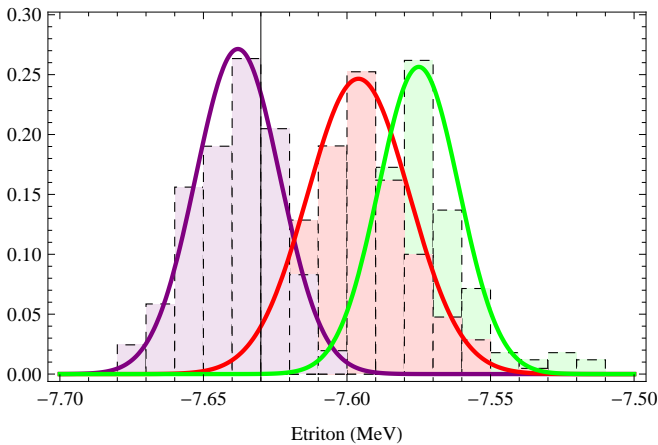}
\caption{(Color online) Normalized histograms representing the triton
  binding energy (in MeV) for a sample of $205$ gaussian potential
  parameters. From left to right the normal density probability
  distribution function $N(\mu,\sigma)$ correspond to higher accuracy
  $E_t= -7.638 \pm 0.0147 $, intermediate accuracy $E_t= -7.596 \pm
  0.0178$ and lower accuracy $E_t= - 7.596 \pm 0.0178 $ (see main text).}
\label{fig:triton}       
\end{figure}

The results of $B_t$ for the 205 Monte Carlo potential samples are
summarized in the histogram of Fig.~\ref{fig:triton} for the three
cases $(i)$, $(ii)$ and $(iii)$ outlined above. As we see, the
propagated histograms are roughly gaussians, with quite similar widths
but shifted. For the most accurate case we get $E_t=-7.638(15) \, {\rm
  MeV}$.  Taking into account the slight asymmetry in the distribution
a $\pm 1\sigma$ ($=68\%$) confidence interval can be obtained by
excluding the $16 \%$ upper and lower tails. This gives the $68\%$
range $ \min E_t \le E_t \le \max E_t $ which corresponds to $\Delta
E_t^{\rm stat} \equiv (\max E_t - \min E_t )/2$.  There is an
uncertainty coming from the fact that for $M=205$ we may exclude 32 or
33 values from above or below, so that
\begin{eqnarray}
\Delta E_t^{\rm stat}= 15(1) \, {\rm keV} \, . 
\end{eqnarray}
This is our main result, which sets a realistic precision for triton
binding energy calculations and is more than two orders of magnitude
larger than the experimental precision $\Delta E_t^{\rm exp}= 0.1 {\rm
  keV}$.  The early estimate $\Delta E_t^{\rm th} > 40 {\rm
  keV}$~\cite{adam1993error} was based on the $^1S_0$ inverse
scattering analysis using the 1980 Paris potential which has a large
$\chi^2/\nu \sim 2$.

It is worth noting that the numerical error in the present calculation
is $\Delta B_t^{\rm num}=1 \, {\rm keV}$ which is one order of magnitude
smaller than $\Delta B_t^{\rm stat}= 15 \, {\rm keV}$. Given that the
error is dominated by the uncertainty of the input potential, we
investigated if the numerical precision can be relaxed, thus reducing
the computing time. Obviously, the meaning of numerical precision may
depend on the method and different approaches should be tried out. The
convergence of the binding energy calculation in terms of partial
waves (see e.g. Ref.~\cite{wu1993three} for explicit notations) is
presented in Table~\ref{tab:Etriton} for one potential taken at random
and whose total energy is given by $E_t=-7.6510 \, {\rm MeV}$. There, an
increasing number of channels is added depending on the relative
orbital angular momenta $(L,l)$ of a NN pair or the third spectator
nucleon respectively (denoted as $(l_x,l_y)$ in the HH expansion above).  As
one can see one needs the $Ss, Sd, Ds$ channels to get a bound
triton $E_t=-7.0117 \, {\rm MeV}$. Within this reduced Hilbert space we
get
\begin{eqnarray}
\Delta B_t^{\rm stat}(Ss+Sd+Ds)= 20 \, {\rm keV}
\end{eqnarray}
When the $Pp$ channel is added, we obtain $\Delta B_t^{\rm
  stat}(Ss+Sd+Ds+Pp) = 19 \, {\rm keV}$.  So, about $75\%$ of the
statistical uncertainty comes from the lowest $Ss+Sd+Ds$ channels

\begin{table}[th]
	\centering
      \caption{Triton binding energy convergenge in the number of
        channels, $N_c$, classiffied according to the orbital angular
        momentum of the pair $L_{\rm Pair}$ and the spectator $l_{\rm
          spectator}$ in the triton as the number of total accumulated
        channels, $N_{\rm Total}$, is increased.  The potential used
        was Monte Carlo generated. A horizontal line is drawn when the
        change in $E_t$ is {\it smaller} than the statistical
        uncertainty $\Delta B_t = 15(1) \, {\rm keV}$.}
      \label{tab:Etriton}
	\begin{tabular*}{\columnwidth}{@{\extracolsep{\fill}}c c c c }
            \hline
            \hline\noalign{\smallskip}
             $N_c$ & $L_{\rm Pair} \, l_{\rm Spectator}$ &  $N_{\rm Total}$ & Energy (MeV)\\
            \hline\noalign{\smallskip}
             3 & Ss  & 3 &  Unbound \\ 
             +2  & Sd+Ds & 5 & -7.0117  \\ 
             +10  & Pp    & 15 & -6.4377 \\ 
             +8 & Dd    & 23 & -7.4109  \\ 
             +4  & Pf+Fp & 27 & -7.4956  \\ 
             +10  &  Ff   & 37 & -7.5654  \\ 
             +2  & Dg+Gd  & 39 &   -7.6178 \\ 
             +8  & Gg  &  47 &  -7.6502 \\ \hline 
             +4  & Fh+Hf  & 51 &   -7.6508 \\ 
             +10  & Hh  &  61 &  -7.6510 \\      
            \noalign{\smallskip}\hline
            \hline
	\end{tabular*}
\end{table}

One interesting aspect from the present analysis concerns the
statistical correlation analysis of the NN gaussian potential
parameters, as this helps to pin down what does fix the current
precision. We find that correlations are never larger than $0.4$, but
since the gaussian potential parameters themselves are strongly
correlated there is still the possibility that more global parameters
such as volume integrals or low energy scattering parameters would
show a clearer pattern.

The precision has been a recurrent topic within the present context,
and much of the effort was originally directed with the purpose of
establishing the need of 3N-forces within the numerical precision of
the calculations.  For instance, one needs 34 channels up to angular
momentum $J_{\rm pair} \le 4$ to obtain $\Delta E_t^{\rm num}=10 {\rm
  keV}$~\cite{Chen:1985zzb}. Within this numerical precision the
triton binding energy obtained by Faddeev calculations has been found
to be $ 8.00, 7.62, 7.63, 7.62, 7.72 $ MeV for the CD
Bonn~\cite{Machleidt:1995km}, Nijm-II, Reid93, Nijm-I and
AV18~\cite{Friar:1993kk} respectively. The covariant spectator model
has produced the closest binding energy $8.50$ MeV to experiment
precisely when the NN $\chi^2$ becomes smallest. The spread of values
in $B_t$, allowed by the theorem of Gl\"ockle and
Polyzou~\cite{polyzou1990three}, is coming from off-shell ambiguities.
The theorem however, does not predict quantitatively the dispersion,
which yields $B_t = 7.85(34) \, {\rm MeV}$ (exp. $B_t = 8.4820(1) \,
{\rm MeV}$). The similarity of the databases but the different
potential forms suggests calling this a systematic error, i.e.
$\Delta B_3^{\rm syst} = 340 {\rm keV}$. In previous estimates a value
of $B_t=7.62(1)$ was obtained using the NijmII, AV18 and Reid93 local
potentials fitted to the same database~\cite{Friar:1993kk}.  This was
extrapolated to be $B_t=7.6(1)$~\cite{Gibson:1994sd} from an inverse
scattering analysis of Nijmegen phases up to $T_{\rm LAB}=300 \, {\rm
  MeV}$ based on a local potential, the error stemming from the high
energy extrapolation. We note that these are essentially systematic
error estimates.

A high precision calculation with the AV18 potential using the HH expansion
method was carried out by the Pisa group~\cite{kievsky1997high}
leading to the sequence of values $B_t= 7.59267, 7.61227, 7.61786,
7.61809, 7.61812 { \rm MeV} $ for $N_c=8,14,18,22,26$ channels
respectively. According to our error estimate of $\Delta B_t=0.02 {\rm
  MeV}$ one could stop already at $N_c=8$ for a realistic
precision. Similar remarks apply to~\cite{Elster:1998qv} where $\Delta
B_t^{\rm num}=0.1 \, {\rm keV}$. Based on general arguments, attempts
have also been made to quantify the systematic uncertainties in
nuclear bindings stemming from NN
scattering~\cite{NavarroPerez:2012vr,Perez:2012kt,NavarroPerez:2012qf})
yielding $\Delta E^{\rm sys}/A = 100-500 \, {\rm keV}$ in rough
agreement with the more sophisticated three-body calculations. This
suggests to use the present calculation as a benchmark in approximate
error estimates sidesteping the full fledged calculation.

From a more general perspective, there is an ongoing effort to
quantify the uncertainties in nuclear
physics~\cite{dudek2013predictive,Dobaczewski:2014jga} as a means to
establish the real predictive power of the theory. While this topic is
presently in its infancy, from a theoretical point of view and the
inferred predictive power, errors in {\it ab initio} calculations can
be grouped into three main categories: i) the input information (in
our case the NN scattering experimental data), ii) the method of
solution and its numerical precision and iii) the form (e.g. local or
non-local) of the interaction in the unknown region.  We have denoted
these errors as $\Delta E^{\rm stat}$, $\Delta E^{\rm num}$ and
$\Delta E^{\rm syst}$ respectively. Assuming that these sources of
error are independent of each other we expect the total theoretical
uncertainty to be given by
\begin{eqnarray}
(\Delta E^{\rm th})^2 = (\Delta E^{\rm stat})^2 +(\Delta E^{\rm num})^2 +(\Delta
E^{\rm syst})^2
\end{eqnarray}
Clearly, the total error is dominated by the largest one. So, it makes
sense either to reduce the largest source of uncertainty or to tune
all uncertainties to a similar level. This sets a {\it realistic}
limit of predictive power in {\it ab initio} calculations, which we
find to be $ \Delta E_t^{\rm th} \ge 15(1) \, {\rm keV} $. While the use of
realistic potentials has been a must in few body calculations, we note
that the physical precision of the calculation is finite and will
definitely have sizeable consequences in large scale calculations in
nuclear physics.  Given the large systematic uncertainties, the
theoretical calculation of the triton binding energy provides a good
example of a precise but inaccurate quantity.

We thank Andreas Nogga for discussions and communications and for
drawing our attention to Ref.~\cite{adam1993error}. This work is
supported by Spanish DGI (grants FIS2011-23565 and FIS2011-24149) and
Junta de Andaluc{\'{\i}a} (grant FQM225).  R.N.P. is supported by a
Mexican CONACYT grant.


\end{document}